\documentclass[11pt,letter]{article}

\usepackage[letterpaper,top=.13in,bottom=.17in,left=.57in,right=.57in,includefoot]{geometry}
\usepackage[natbib=true,style=nature,sorting=none]{biblatex} %Imports biblatex package
\usepackage{graphicx}
\usepackage{booktabs}
\usepackage{fancyhdr}
\usepackage[fleqn]{amsmath}
\usepackage{epstopdf}
\usepackage{amsfonts}
\usepackage{setspace}
\usepackage{bbm}
\usepackage{verbatim}
\usepackage{rotating}
\usepackage{multirow}
\usepackage{txfonts}
\usepackage{blindtext}
\usepackage{multicol}
\usepackage{caption}
\AtBeginEnvironment{tabular}{\small}
\usepackage{enumitem}
\setlist[itemize]{leftmargin=*,topsep=0pt,itemsep=-1ex,partopsep=1ex,parsep=1ex}

\usepackage[compact]{titlesec}
\titleformat*{\section}{\MakeUppercase}

\titleformat*{\subsection}{\it}
\titleformat*{\subsubsection}{\it}

\begin{document}\pagestyle{empty}
\parbox[][1.3in][t]{\textwidth}{%
  {\LARGE Bridging Structural Dynamics and Biomechanics: Human Motion Estimation through Footstep-Induced Floor Vibrations\\~\\} 
  {\Large Yiwen Dong$^{1,2}$,  Jessica Rose$^2$ \& Hae Young Noh$^1$}\\
  \emph{\large $^1$ Civil and Environmental Engineering, Stanford University\\
  $^2$ Department of Orthopaedic Surgery, Stanford Medicine}\medskip
}

%%%%%%%%%%%%%%%%%%%%%%%%%%%%%%
\parbox[][2.9in][t]{\textwidth}{%
%This study aims to estimate lower-limb joint motion during walking through gait-induced floor vibrations. 
ABSTRACT: Quantitative estimation of human joint motion in daily living spaces is essential for early detection and rehabilitation tracking of neuromusculoskeletal disorders (e.g., Parkinson's) and mitigating trip and fall risks for older adults. Existing approaches involve monitoring devices such as cameras, wearables, and pressure mats, but have operational constraints such as direct line-of-sight, carrying devices, and dense deployment. To overcome these limitations, we leverage gait-induced floor vibration to estimate lower-limb joint motion (e.g., ankle, knee, and hip flexion angles), allowing non-intrusive and contactless gait health monitoring in people's living spaces. To overcome the high uncertainty in lower-limb movement given the limited information provided by the gait-induced floor vibrations, we formulate a physics-informed graph to integrate domain knowledge of gait biomechanics and structural dynamics into the model. Specifically, different types of nodes represent heterogeneous information from joint motions and floor vibrations; Their connecting edges represent the physiological relationships between joints and forces governed by gait biomechanics, as well as the relationships between forces and floor responses governed by the structural dynamics. As a result, our model poses physical constraints to reduce uncertainty while allowing information sharing between the body and the floor to make more accurate predictions. We evaluate our approach with 20 participants through a real-world walking experiment. We achieved an average of 3.7 degrees of mean absolute error in estimating 12 joint flexion angles (38\% error reduction from baseline), which is comparable to the performance of cameras and wearables in current medical practices.

%To overcome this challenge, we integrate 1) gait biomechanics, where each combination of joint motion angles result in unique footstep forces, and 2) structural dynamics, where these unique footstep forces generate distinctive dynamic floor responses. Specifically, we develop a physics-informed graph transformer model that represents 1) the physiological relationships among lower-limb joints and 2) the dynamics between footstep force and floor vibration through heterogeneous nodes and edges, allowing heterogeneous information sharing between joint motion and floor vibration. We evaluate our approach with 20 participants through a real-world walking experiment. Results show that our model has an average of 3.7 degrees of mean absolute error in estimating 12 joint flexion angles (38\% error reduction from a Long Short-Term Memory (LSTM) baseline), which is comparable to the performance of cameras and wearables in current medical practices.
}

\begin{multicols*}{2}

%%%%%%%%%%%%%%%%%%%%%%%%%%%%%%
\section{Introduction}
Quantitative estimation of joint motion during walking is essential for clinical detection and rehabilitation of gait disorders such as diabetes and Parkinson's and mitigation of trip and/or fall risks for older adults~\cite{schulte1993quantitative}. Existing approaches for joint motion estimation involve sensing devices such as motion capture (MoCap) systems, video cameras, wearables, and pressure/force sensors, but have operational constraints when used outside the laboratory settings. MoCap systems are commonly used in clinical usage~\cite{pfister2014comparative}, but they require marker installment and dedicated calibration. Pressure/force sensors capture ground reaction forces~\cite{stauffer1977force}, but they lack body motion information. Video cameras and wearables capture body motion~\cite{aggarwal1999human,tao2012gait}, but they can raise privacy concerns and/or cause discomfort when carrying devices all the time. 

In this study, we leverage human gait-induced floor vibration to infer the lower-limb joint motion, which has the benefits of being non-intrusive, wide-ranged, and contactless. The intuition of this approach is that each joint motion combination from the ankle, knee, and hip exerts a unique footstep force to the floor, which generates a unique floor vibration pattern. We capture these vibrations using geophone sensors mounted on the floor surfaces to infer the joint motion combination, allowing gait health monitoring in people's living spaces.

However, the main challenge is the high uncertainty in joint motion given the limited and indirect measurement provided by floor vibrations.
%, making it challenging for conventional data-driven models to infer the ankle, knee, and hip joint angles accurately. 
To overcome this challenge, we represent the indirect relationship between the floor vibration and the joint motions through a physics-informed graphical model. The model integrates 1) gait biomechanics, which describes the physiological relationships among joint motions through the connections of muscles and bones, and 2) structural dynamics, which describes the dynamic floor responses under footsteps governed by the structural dynamics equations. In our graph, different types of nodes represent heterogeneous information from joint motions and floor vibrations; Their connecting edges represent the physiological relationship between joints and forces, as well as the dynamics between forces and floor responses. The formulation of the graph poses physical constraints while allowing information sharing among heterogeneous data. 

The contributions of the study are that we:

\begin{enumerate}
    \item Develop a novel approach to estimate lower-limb joint motion for gait health monitoring using footstep-induced floor vibrations;

    \item Integrate structural dynamics and gait biomechanics to formulate a new human-structure interaction system and develop a physics-informed graphical model to reduce uncertainties in joint motion estimation for gait health monitoring;

    \item Evaluate our approach through a real-world experiment and obtain promising results.
\end{enumerate}

We evaluated our approach with 20 participants for 4 gait types commonly observed in clinics, through collaboration with the Lucile Packard Children's Hospital at Stanford. 
%Our approach achieved an average of 3.70 degrees of mean absolute error in estimating 12 critical joint flexion angles, which has reduced $38\%$ of the error from the baseline LSTM model (6.02 degrees). 
The accuracy is comparable to other sensing approaches used in medical practices such as cameras and wearables~\cite{majumder2020wearable,finkbiner2017video}.

\section{Bridging Structural Dynamics and Gait Biomechanics}\label{sec:2}
We formulate a new human-structure interaction (HSI) system by integrating structural dynamics and gait biomechanics through their common connection with the ground reaction forces (see Figure~\ref{fig:newHSI}). Our formulation closes the gap in existing work by inferring human posture for health~\cite{dong2024graphical}.

\includegraphics[width=3.53in]{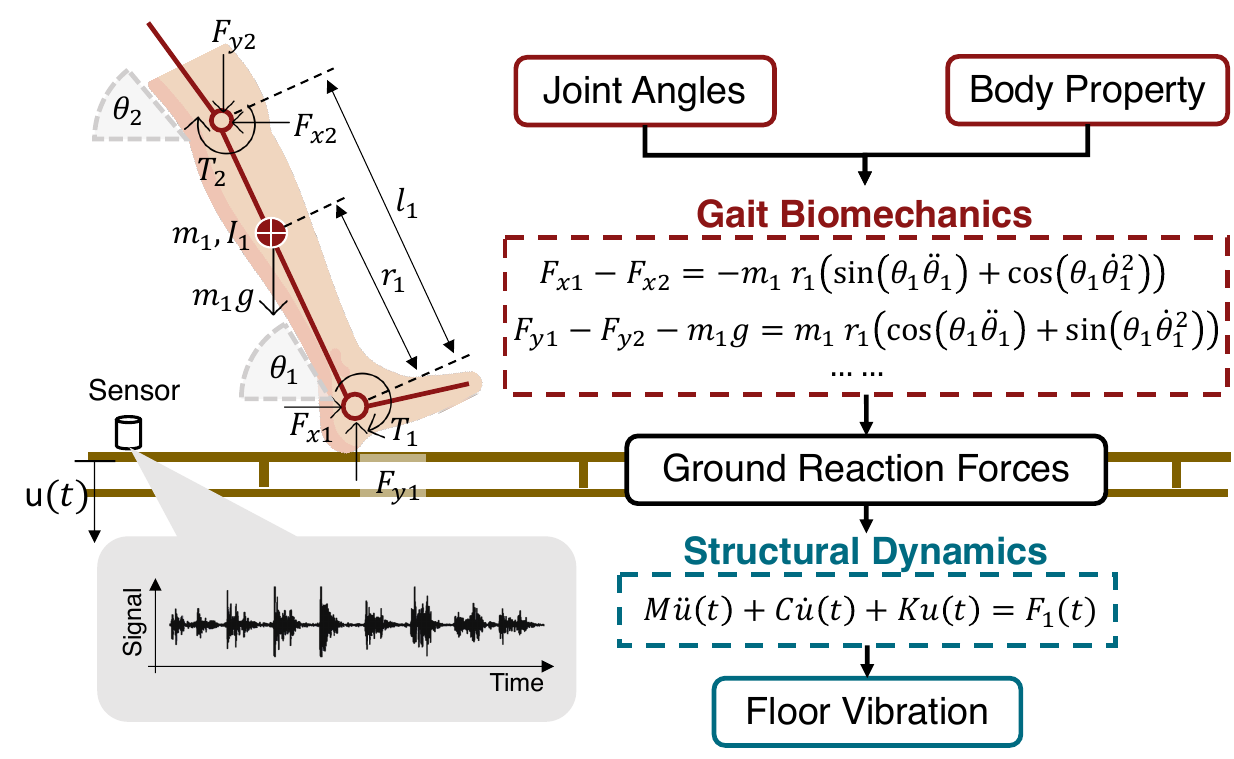}
\captionof{figure}{Conceptual diagram of our new HSI system.}\label{fig:newHSI}

\vspace{0.1in}

% \subsection{Bridge between Gait Biomechanics and Structural Dynamics using Ground Reaction Forces}
%We develop a new formulation of the HSI system to infer human gait health from floor vibration. While there are existing HSI systems to assess the floor vibration under pedestrian loading for serviceability requirements, they do not consider the posture of the person, making them unable to assess gait health during walking. Our formulation integrates gait biomechanics with structural dynamics to bridge the gap between structural vibration sensing and gait health monitoring, illustrated in Figure~\ref{fig:newHSI}.

The physical insight of our HSI system can be divided into two parts. First, the human gait (represented by joint angles and body properties) exerts forces onto the floor, which is governed by gait biomechanics; Then, the ground reaction forces induce floor vibrations, which are affected by the floor properties and governed by structural dynamics equations. The details of these two parts are described below. %Our system bridges gait biomechanics and structural dynamics to establish the relationship between joint motion and floor vibration.

\textbf{Gait Biomechanics.} The inverse dynamics in gait biomechanics describe the relationship between joint angles and ground reaction forces, where each section of the leg is analyzed through a free-body diagram with forces and moments applied. The upper part of Figure~\ref{fig:newHSI} shows an example of the biomechanics of the shank. The complete analysis involves thigh, shank, and foot sections, where the hip, knee, and ankle joints are regarded as hinges for force/moment transmission. The above equations assume the body is symmetrical, the foot has negligible mass, and the motion in the frontal plane is negligible.

\textbf{Structural Dynamics.} The dynamics of the floor structures under footstep forces are typically represented through the equation of motion, as shown in the lower part of Figure~\ref{fig:newHSI}. The equation suggests that, as the foot exerts forces $F_1(t)$ to the floor, the resultant vibration which depends on the mass, stiffness, and damping of the floor, is captured by the sensor mounted on the floor surface as velocity $\dot{u}(t)$ or acceleration $\ddot{u}(t)$.

Our HSI formulation forms a complete chain of physical relations from human gait to floor vibrations, allowing formal analysis and inference of gait health information.

\begin{figure*}
\includegraphics[width=7in]{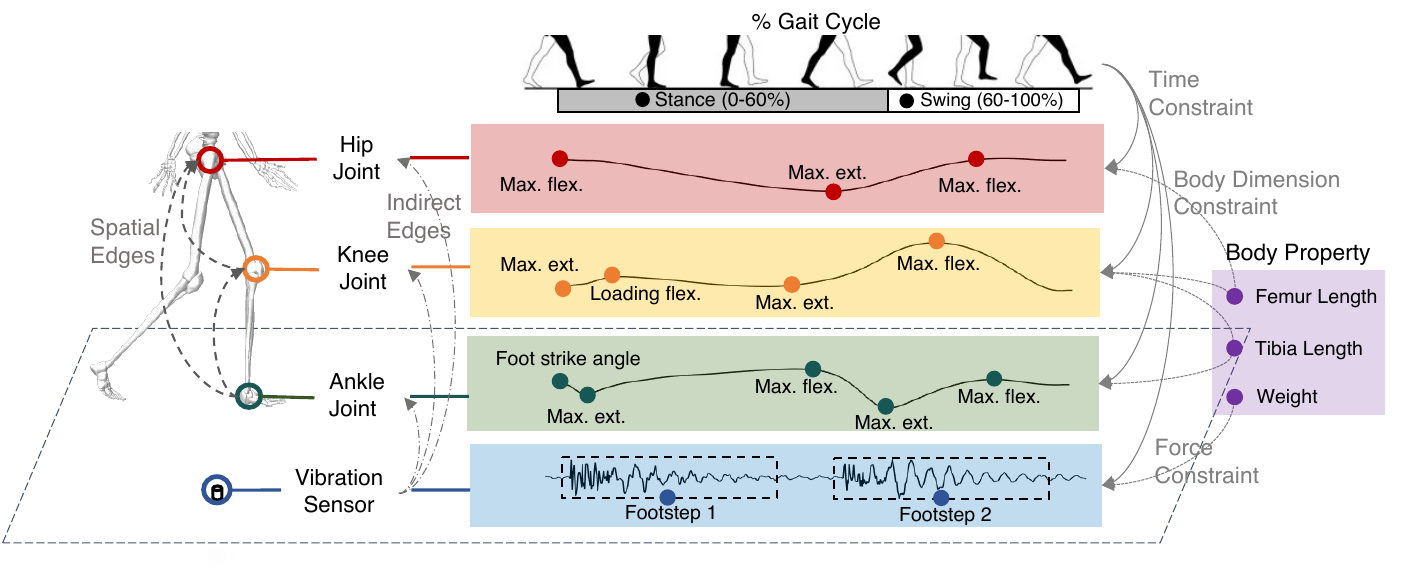} 
\captionof{figure}{Diagram of our physics-informed graph (PIG) describing the relationships between hip, knee, ankle joint motions, floor vibrations, gait cycle time, and body properties. The nodes of the graph are represented as solid circles with various colors. The edges are represented as arrows. %Converting the conceptual model (right) to the physics-informed heterogeneous graph (left) to describe the relationships between hip, knee, and ankle joints and floor vibrations.
}\label{fig:graph}
\end{figure*}

\section{Joint Motion Estimation through Physics-Informed Graphical Model}

To infer joint motion from floor vibration for gait health assessments, we develop a physics-informed graphical (PIG) model based on our HSI formulation. We first model the gait and floor information through a heterogeneous graph with nodes and edges, and then design the information flow in the graph to pose physical constraints during model training to reduce uncertainty.

\subsection{Modeling Gait and Floor Information}
The graph consists of 4 types of nodes to represent the gait and floor information and 5 types of edges to model the physiological and structural dynamics relationships. The nodes are depicted as dots in Figure~\ref{fig:graph}, summarized below:
\begin{itemize}
    \item \textbf{Joint Nodes:} There are three types of joint nodes, each representing the hip, knee, and ankle joint motion. The information stored in each joint node is the magnitude of critical flexion/extension angles over each gait cycle, as described in Figure~\ref{fig:graph}. These are chosen because they are important indicators of gait abnormalities and inform potential trip/fall risks for doctors in gait clinics.
    \item \textbf{Time Nodes:} The time nodes contain important moments when critical joint angles happen in a gait cycle, including the foot strike time and foot off time that divides a gait cycle into the stance and swing phases.
    \item \textbf{Vibration Nodes:} The vibration nodes store the vibration generated on the floor recorded by sensors, which contains a vibration signal segment based on the beginning and the end of a gait cycle.
    \item \textbf{Body Nodes:} The body nodes describe the anthropometry of the walker, such as the body weight and leg lengths. These are important variables to determine the ground reaction forces, as discussed in Section~\ref{sec:2}.
\end{itemize}
\vspace{0.1in}

The relationships between various nodes are defined by edges. Figure~\ref{fig:graph} summarizes these relationships by presenting the relative position of joints and sensors in space (vertical direction), as well as the critical joint motions and vibration data in time (horizontal direction). These relationships are represented by various types of edges:
\begin{itemize}
    \item \textbf{Spatial Edges:} The spatial edges represent the physiological connection among hip, knee, and ankle joints. Since the change of motion in one joint affects the others as they are connected through muscles and bones, the spatial edges model such dependency and allow information sharing among various joints.
    \item \textbf{Temporal Edges:} The temporal edges connect within the same type of joint nodes, which describes the sequence of motion over a gait cycle such as the knee extension at the footstrike, flexion at loading time, extension at the foot off, and flexion during the swing phase~\cite{dong2024graphical}. 
    \item \textbf{Indirect Edges:} The indirect edges refer to the connection between the joint and vibration nodes. These edges encode the indirect relationship between the joint motion and the vibration data, which allows joint motion estimation through dynamic floor responses.
    \item \textbf{Time Constraint Edges:} The time constraint edges connect the time nodes with the joint nodes, representing the relationship between joint motion and the gait cycle, as described in our prior work~\cite{dong2023structure,dong2024ubiquitous}.
    \item \textbf{Body Dimension Constraint Edges:} These edges connect the joint nodes with body nodes that describe lower-limb lengths, allowing joint forces and moments to be estimated through gait biomechanics.
    \item \textbf{Force Constraint Edges:} This constraint bridges body weight and floor vibration through the ground reaction force. The main insight is that the ground reaction force is typically proportional to the body weight, resulting in larger vibration amplitudes.
\end{itemize}

The main benefit of this physics-informed graph is to incorporate complex dependencies and integrate heterogeneous information over time and space. By establishing the relationship among joints, vibrations, body properties, and gait cycle time, our model systematically reduces uncertainties for gait health monitoring.

\subsection{Posing Physical Constraints to Reduce Uncertainty}
In this section, we introduce the physics-informed graph (PIG) model, which allows training on a physics-informed graph for joint angle estimation. One main challenge in model training is the data requirement - due to the high complexity of graph formulation, it requires a large amount of walking data from each person, which is not practical for people with walking impairments. To overcome this challenge, %our model actively controls the information flows along the edges of our graph by enforcing physical equations in structural dynamics and gait biomechanics. 
we pose physical constraints to reduce uncertainty in data by enforcing equations in structural dynamics and gait biomechanics to control the data flow along the edges of our graph. This improves data efficiency and ``teaches'' our model to follow physical laws.

\includegraphics[width=3.53in]{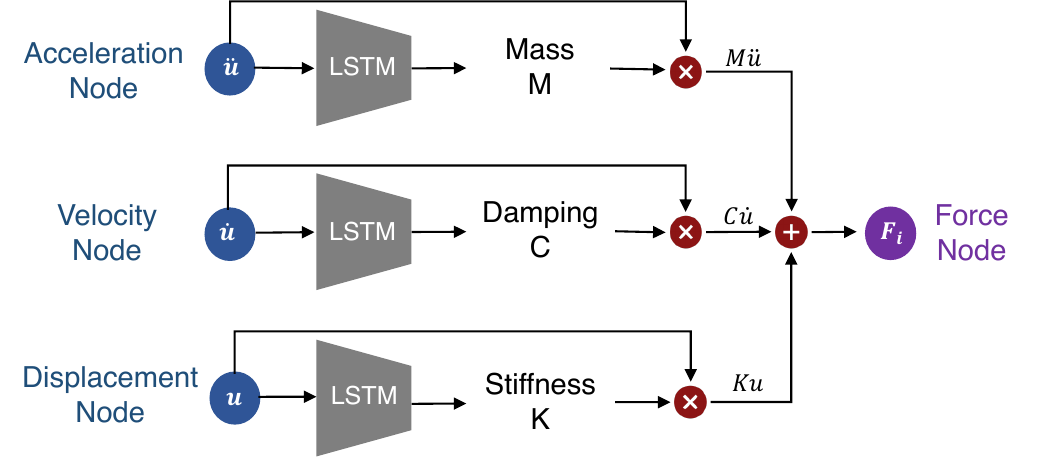}
\captionof{figure}{Information flow between the vibration nodes and force nodes to enforce structural dynamics equation.
}\label{fig:dynamics_flow}
% \subsubsection{Enforcing Structural Dynamics Through Multi-Node Aggregation}\label{sec:structmodel}
\vspace{0.1in}

First, we enforce structural dynamics by controlling how information aggregates from various vibration nodes to the force nodes, illustrated in Figure~\ref{fig:dynamics_flow}. Given that the relationship between ground reaction force $F_i(t)$ and floor vibration $u(t)$ is governed by the equation of motion $Mu(t) + C\dot{u}(t) + K\ddot{u}(t) = F_i(t)$, we first use the summation as our aggregation function. Then, we infer the typically unknown structural properties (mass, damping, and stiffness of the floor) in practical scenarios by developing ``structure property learners'' (represented as the long-short term memory (LSTM) modules in Figure~\ref{fig:dynamics_flow}) to implicitly extract the structural information. This controls information flow and enforces structural dynamics in our PIG model.

% \subsubsection{Modeling Gait Biomechanics Through Node-Level Transformation}\label{sec:biomodel}
Similarly, we enforce the gait biomechanics equations to reduce uncertainties in joint nodes. First, we transform the features at each joint to a space defined by $sin$ and $cos$ and body dimensions to the weights of leg sections and lengths of moment arms to approximate the biomechanics equation in Figure~\ref{fig:newHSI}. Then, we leverage the attention mechanism~\cite{brody2021attentive} to determine the importance of each transformed feature and create multiplied terms between joint and body nodes. Finally, we aggregate the information among the multiplied terms to pass on to the force node. Combining all the steps above, we enforce gait biomechanics in our PIG model.

\section{Real-World Evaluation with Commonly Observed Gait Types}
We evaluate our approach through a 20-subject walking experiment by collaborating with medical doctors. The experiment setup and results are discussed in this section.

\subsection{Experiment Setup}
During the experiment, four commonly observed gait types are tested, including normal gait, toe-walking, flexed-knee, and gait with foot drag (see Figure~\ref{fig:exp}). Each participant first walks with their healthy natural gait for 20 trials, and then with ``simulated'' abnormal gait under the instruction of gait experts from Lucile Packard Children's Hospital for 10 trials each. All experiments are conducted following the approved IRB (IRB-55372).

\includegraphics[width=3.5in]{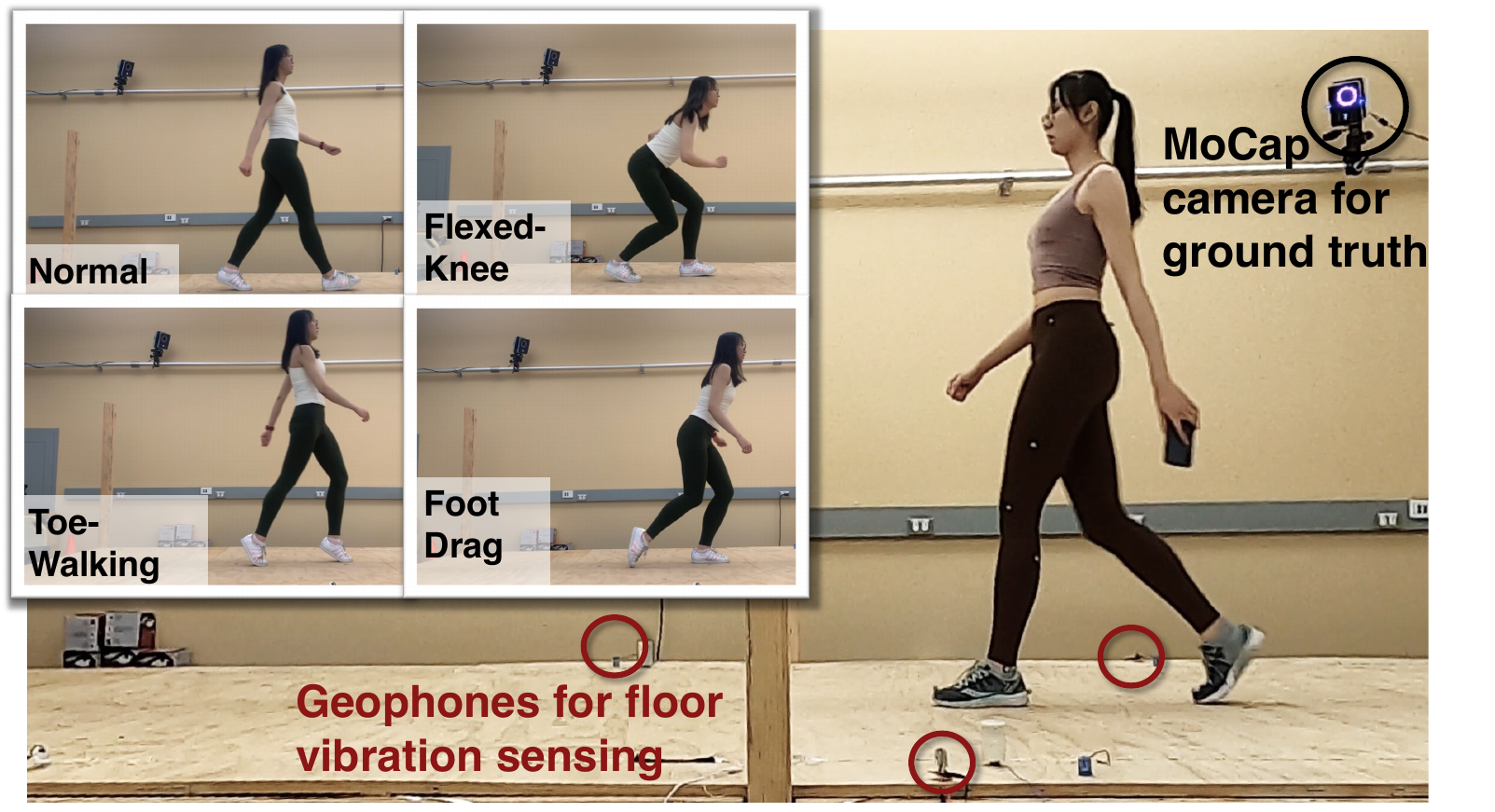}
\captionof{figure}{Experiment setup with geophones and Motion Capture (MoCap) cameras for real-world walking.
}\label{fig:exp}
\vspace{0.1in}

The vibration data collection system includes four SM-24 geophone sensors mounted on the floor surface with a 500 Hz sampling frequency. The signals from the sensors are amplified by 500$\times$ to improve the signal-to-noise ratio. The ground truth joint motions are captured by a Vicon MoCap System with a frame rate of 100 fps. During the experiment, ten infrared cameras recorded 3-dimensional trajectories of lower-limb joints when walking. The joint angles are computed by the Vicon Plug-in Gait lower body model. A Vicon Lock Lab system is used to synchronize the vibration data with the lower-limb motion data.

\subsection{Results and Discussion}
Our approach has an average of 3.7 degrees of mean absolute error (MAE) in estimating 12 joint flexion angles on test data, which significantly outperforms the existing baseline (38\% error reduction) with 5.1 degrees MAE by using a fine-tuned LSTM model. The results breakdown for each motion segment is shown in Figure~\ref{fig:results}.

\includegraphics[width=3.53in]{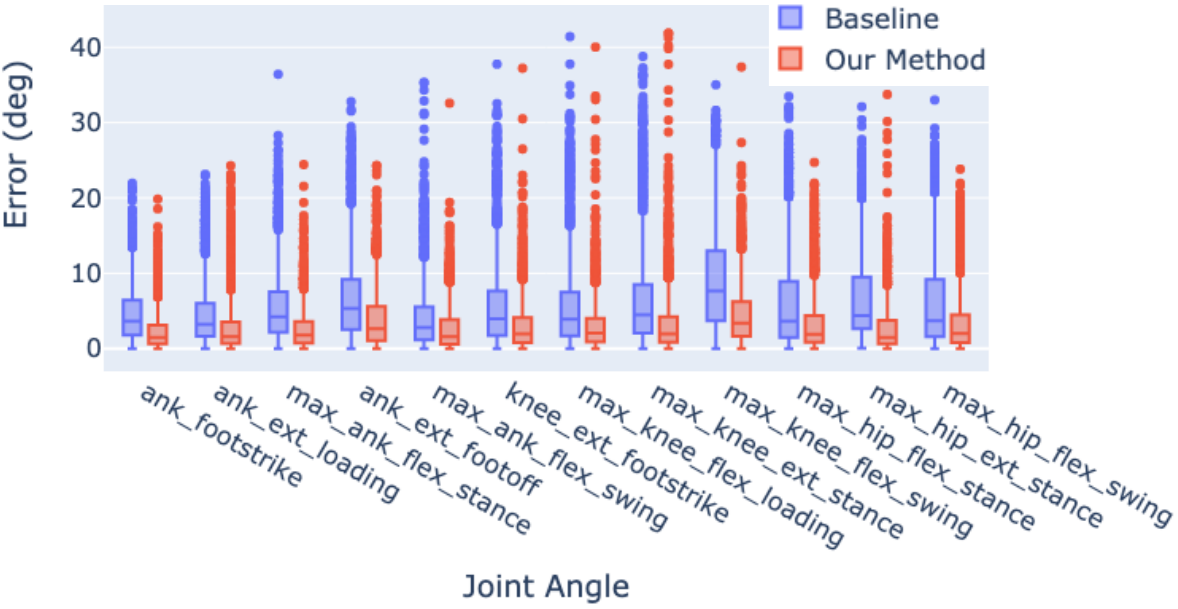}
\captionof{figure}{Results comparison between baseline model (LSTM) and our physics-informed graphical (PIG) model. 
}\label{fig:results}

\subsubsection{Comparison among Various Joints.} Among all the joint motion segments, our method has a relatively lower error on the ankle joint angles. This may be because the ankle motion directly influences the interaction between the foot and the floor, making it easier to infer from floor vibration. On the other hand, the swing phase joint angles have higher errors than the stance phase. This may be due to the lack of contact between the foot and the floor when the leg swings in the air. 

\subsubsection{Comparison among Various Gait Types}
Our results in joint angle estimation for normal walking (1.7 degree MAE) is significantly lower than that of the abnormal gait (4.4 degree MAE). This is because normal walking patterns are relatively consistent among various trials while the abnormal gait has significantly higher variance due to the instability of the posture.

\subsubsection{Comparison with Various Sensing Devices.} Our approach has comparable accuracy with the existing state-of-the-art sensing devices. For example, Majumder et al. evaluated the on-body wearables and reported a 2 to 3.4 degree of RMSE error~\cite{majumder2020wearable}. Finkbiner et al. developed video-based human pose estimation and reported 3.1 to 5.8 degrees of MAE for estimating joint angles~\cite{finkbiner2017video}. By comparison, our approach (3.7 degrees MAE) has a similar scale of accuracy while providing a non-intrusive and device-free user experience.

\section{Conclusion}
In this study, we estimate joint motions using footstep-induced floor vibrations by integrating structural dynamics and gait biomechanics. To overcome the high uncertainty challenge, we develop a physics-informed graphical model to enforce structural dynamics and gait biomechanics equations. Through a walking experiment with 20 people, we obtained 3.7 degrees of MAE in joint angle estimation, which is comparable to the existing portable devices.

\section*{References}
{\fontsize{9pt}{11pt}\selectfont
[1] Schulte, L., et al. (1993). A quantitative assessment of limited joint mobility in patients with diabetes. Arthritis \& Rheumatism: Of-ficial Journal of the American College of Rheumatology, 36(10), 1429-1443.\par
[2] Pfister, A., West, A. M., Bronner, S., \& Noah, J. A. (2014). Compara-tive abilities of Microsoft Kinect and Vicon 3D motion capture for gait analysis. Journal of medical engineering \& technology, 38(5), 274-280.\par
[3] Stauffer, R. N., Chao, E. Y., \& Brewster, R. C. (1977). Force and mo-tion analysis of the normal, diseased, and prosthetic ankle joint.  Clinical Orthopaedics and Related Research (1976-2007), 127, 189-196.\par
[4] Tao, W., Liu, T., Zheng, R., \& Feng, H. (2012). Gait analysis using wearable sensors. Sensors, 12(2), 2255-2283.\par
[5] Aggarwal, J. K., \& Cai, Q. (1999). Human motion analysis: A review. Computer vision and image understanding, 73(3), 428-440.\par
[6] Finkbiner, M. J., et al. (2017). Video movement analysis using smart- phones (ViMAS): a pilot study. Journal of Visualized Experiments, 121, e54659.\par
[7] Majumder, S., \& Deen, M. J. (2020). Wearable IMU-based system for  real-time monitoring of lower-limb joints. IEEE sensors journal, 21(6), 8267-8275.\par
[8] Dong, Y., \& Noh, H. Y. (2023). Structure-Agnostic Gait Cycle Segmen-tation for In-Home Gait Health Monitoring Through Footstep-Induced Structural Vibrations. Society for Experimental Mechan-ics Annual Conference and Exposition (65-74). Springer Nature. \par
[9] Dong, Y., \& Noh, H. Y. (2024). Ubiquitous gait analysis through footstep-induced floor vibrations. Sensors, 24(8), 2496. \par
[10] Dong, Y., Liu, J., Kim, S. E., Schadl, K., Rose, J., \& Noh, H. Y. (2024). Graphical Modeling of the Lower-Limb Joint Motion from the Dynamic Floor Responses Under Footstep Forces. In IMAC, A Conference and Exposition on Structural Dynamics (pp. 9-16). Springer Nature Switzerland. \par
[11] Brody, S., Alon, U., \& Yahav, E. (2021). How attentive are graph atten-tion networks? arXiv preprint arXiv:2105.14491.\par
}

\end{multicols*}

\end{document}